\documentstyle[12pt]{article}

\begin{document}
\begin{titlepage}

\begin{flushright}
IFUP--TH--56/95
\end{flushright}
\vskip 1truecm
\begin{center}
\Large\bf
Conformal gauge fixing and Faddeev--Popov \\
determinant in 2-dimensional Regge gravity\footnote{This work is
  supported in part by M.U.R.S.T.}
\end{center}
\vskip 1truecm
\begin{center}
{Pietro Menotti and Pier Paolo Peirano} \\
{\small\it Dipartimento di Fisica dell'Universit\`a, Pisa 56100,
Italy and}\\
{\small\it INFN, Sezione di Pisa}\\
\end{center}
\vskip .8truecm
\begin{center}
October, 1995
\end{center}
\vskip 2cm
\begin{center}
Talk given at the XVIII International Workshop \\
on High Energy Physics and Field Theory.\\
Protvino, Russia, June 1995.
\end{center}

\end{titlepage}

\begin{abstract}
By regularizing the conical singularities by means of a segment of a
sphere or pseudosphere and then taking the regulator to zero, we
compute exactly the Faddeev--Popov determinant related to the
conformal gauge fixing for a piece-wise flat surface with the topology of
the sphere. The result is analytic in the opening angles of the
conical singularities in the interval ($\pi$, $4\pi$) and in the
smooth limit goes over to the continuum expression.
The Riemann-Roch relation on the dimensions of ker$(L^{\dag}L)$ and
ker$(LL^{\dag})$ is satisfied.
\end{abstract}

At present we have a well developed and consistent theory of two dimensional
gravity in the continuum formulation \cite{poly,alva} accompanied
by a collection of exact results \cite{ginspa}.

In order to extend the predictive range of such a theory, discretized
versions have been put forward, which can be subject to numerical
simulations. In practice two methods have been proposed and exploited:
the first is the Regge approach \cite{regge,hamb,lee} and the second
the so called dynamically triangulated random surfaces approach
\cite{amb} which in two dimensions has strong connections with the
matrix models \cite{ginspa2}.

Here we shall be concerned with the Regge approach which was historically
the first. Since the beginning the main discussion with regard to the
quantum formulation of Regge gravity centered about the integration
measure to be adopted in the functional integral and the role played
by the diffeomorphisms \cite{lee,diff,jev}.

Here we  shall adhere to the viewpoint that the only
difference between Regge gravity and the usual formulation of quantum
gravity, is that while in the continuum one integrates (or try to
integrate) over all surfaces, in the Regge approach one limits oneself
 to integrate only on the piecewise flat surfaces. When the
number of faces goes to infinity one hopes to recover the continuum
theory.

{}From this viewpoint there is no difference in the definition of
diffeomorphism between the continuum and the Regge formulation
\cite{lee,jev,sork}. A strictly related point is that of the
integration measure. The usual procedure which has proven successful
in gauge theories is to start from the integration over to the local
gauge variables; if the gauge volume turns out to be infinite, a gauge
fixing has to be introduced and the corresponding Faddeev-Popov
determinant computed. The last step is avoided in the lattice
formulation of QCD as the local integration variables are taken as the
elements of the gauge group which is compact and possesses a finite
volume.

In gravity we shall stick to integrating over the analogous of the
gauge variables i.e. the metric. In order to provide a measure, a
distance functional has to be introduced and we shall adopt the
De-Witt metric
\begin{equation}
(\delta g^{(1)}, \delta g^{(2)}) = \int \! \sqrt{g} \; \delta
   g^{(1)}_{\mu\nu} \left( g^{\mu\alpha}g^{\nu\beta} +
   g^{\mu\beta}g^{\nu\alpha} + C g^{\mu\nu}g^{\alpha\beta} \right)
   \delta g^{(2)}_{\alpha\beta},
\end{equation}
which is the only ultra--local metric invariant under diffeomorphisms.

Due to the infinite volume of the diffeomorphisms a gauge fixing has
to be introduced and the most suitable one in two dimensions appears
to be the conformal gauge fixing $g_{\mu\nu} = e^{2\sigma}
\hat{g}_{\mu\nu}(\tau_{i})$, where $\tau_{i}$ are the Teichm\"uller
parameters. In a classical series of papers \cite{poly,alva} the
following expression was reached for the partition function of two
dimensional euclidean quantum gravity
\begin{equation}
{\cal{Z}} = \int {\cal{D}} [\sigma] \, d\tau_{i} \;
\sqrt{\frac{{\det}'(L^{\dag}L)}{\det(\Psi_{k}, \Psi_{l})
    \det(\Phi_{a},\Phi_{b})}}
\label{part}
\end{equation}
where
\begin{eqnarray}
(L\xi)_{\mu\nu} = \nabla_{\mu}\xi_{\nu} + \nabla_{\mu}\xi_{\nu}
- g_{\mu\nu} \nabla^{\rho}\xi_{\rho} \nonumber \\
(L^{\dag}h)_{\nu} = -4 \nabla^{\mu} h_{\mu\nu}
\end{eqnarray}
being $\xi_{\mu}$ a vector field and $h_{\mu\nu}$ a symmetric
traceless tensor field.  $\cal{D}[\sigma]$ is the functional
integration measure induced by the distance
\begin{equation}
(\delta \sigma^{(1)}, \delta \sigma^{(2)}) =  \int \sqrt{\hat{g}}~
e^{2\sigma} \delta \sigma^{(1)} \delta \sigma^{(2)}.
\end{equation}
$\Psi_{k}$ and $\Phi_{a}$ are respectively the zero modes of
$L$ and $L^{\dag}$. For the sphere topology, to which we shall refer
from now on, there are no Teichm\"uller parameters and hence no zero
modes of $L^{\dag}$.
The dependence on $\sigma$ of the integrand in (\ref{part}) can be
factorized in the expression $e^{-26 S_{L}}$ where
\begin{equation}
S_{L}[\sigma,\hat{g}(\tau_{i})] = \frac{1}{24\pi} \int d^2 x \,
\sqrt{\hat{g}} \; [ \hat{g}^{\mu\nu} \partial_{\mu} \sigma
\partial_{\nu} \sigma + R_{\hat{g}} \sigma ] \label{liouv}
\end{equation}
$e^{-26 S_{L}}$ is the (square-root) of the Faddeev-Popov determinant
related to the conformal gauge fixing.

A Regge surface whose singularities have location $\omega_{i}$ in
the projective plane and angular aperture $2\pi\alpha_{i}$
($\alpha_{i}=1$ is the plane), is described by a conformal factor
\cite{foer} $e^{2\sigma} = e^{2\lambda_{0}} \prod_{i} | \omega
-\omega_{i}|^{2(\alpha_{i} -1)}$ which in the neighborhood of
$\omega_{i}$ becomes $e^{2\lambda_{i}} |\omega -
\omega_{i}|^{2(\alpha_{i} -1)}$ with $e^{2\lambda_{i}} =
e^{2\lambda_{0}} \prod_{j \neq i} | \omega_{i} - \omega_{j}
|^{2(\alpha_{j} -1)}$.  In the conformal gauge $L$ and $L^{\dag}$
assume the form
\begin{equation}
L = e^{2\sigma} \frac{\partial}{\partial\bar{\omega}} e^{-2\sigma},
\ \ \ L^{\dag} = - e^{-2\sigma} \frac{\partial}{\partial\omega}.
\end{equation}

If  now we try to evaluate $S_{L}$ for a conformal factor describing a
piecewise flat geometry we obtain a divergent result. Nevertheless
$\displaystyle \frac{{\det}' (L^{\dag}L)}{\det (\Psi_{k}, \Psi_{l})}$
can be defined also for a Regge surface by means of the $Z$--function
regularization \cite{chee} which gives
\begin{equation}
-\ln ({\det}' (L^{\dag}L)) = Z'(s) |_{s=0} =
\gamma_{E} Z(0) +{\mbox{Finite}}_{\epsilon \rightarrow 0}
\int_{\epsilon}^{\infty} \frac{dt}{t} \mbox{Tr}'(e^{-tL^{\dag}L})
\label{zeta}
\end{equation}
where ${\det}'$ and Tr$'$ mean that the zero modes are
excluded. Following the standard procedure developed in the continuum
approach, $Z'(0)$ will be computed by first performing a variation
$\delta\sigma$ in the conformal factor and later integrating back the
result.  For the case at hand we have
\begin{eqnarray}
 \delta Z'(0) = \gamma_E\delta c^{K}_0 +  {\rm Finite_{\epsilon
\rightarrow 0}} \int d^2 x [ 4 \delta
\sigma(x) (K(x,x,\epsilon) -\sum_{k} |\Psi_{k}
(x) |^{2}) +   & \nonumber \\
 - 2 \delta   \sigma(x) H(x,x,\epsilon) ]  & \label{var}
\end{eqnarray}
where $K$ is the heat kernel of the operator $L^{\dag}L$, $H$ is the
heat kernel of the operator $LL^{\dag}$ and $c^{K}_0 = Z(0) + {\rm dim
{}~ker}~(L^{\dag}L)$ is the constant term in the
asymptotic expansion of the trace of $K(x,x',t)$.

Aurell and Salomonson \cite{aur} gave the determinant of the scalar
Laplace-Beltrami operator for a piece-wise flat surface with the
topology of the sphere and for also for some compact domains of the
plane.

In our case \cite{pmppp} the main point is the computation of
$K(x,x',t)$ and $H(x,x',t)$ which respect the correct boundary
conditions imposed by the nature of the vector field $\xi$ and of the
tensor field $h$.  In the neighborhood of $\omega_{i}$ the conical
singularity cone can be described in the $z=x+iy$ plane, by a wedge of
angular opening $2\pi\alpha=2\pi\alpha_{i}$.  The conformal variation
considered here is the same adopted by \cite{aur} for the scalar case,
and it takes from a cone with some opening angle $2\pi \alpha$ and
scale factor $\lambda=\lambda_{i}$ to a cone with varied opening angle
$\alpha +\delta\alpha$ and scale factor $\lambda +\delta\lambda$.
Such a conformal transformation is described by the variation in the
$z$-plane
\begin{equation}
\delta \sigma(z,\bar z) =(\delta \lambda - \lambda
{\delta \alpha\over \alpha})
+ {\delta \alpha\over \alpha} \log(\alpha|z|).
\label{delta}
\end{equation}

The spectral representation of the heat kernel of $L^{\dag}L$ on the
cone is given by
\begin{eqnarray}
K_{\alpha,\delta}(x,x';t) = \frac{1}{2\pi\alpha} \left\{
   \sum_{n=0}^{\infty} e^{i (\phi -\phi') \frac{n+\delta}{\alpha}}
   \int_{0}^{\infty} J_{\frac{n+\delta}{\alpha}}(r\mu)
   J_{\frac{n+\delta}{\alpha}} (r'\mu) e^{-t\mu^{2}} \, \mu d
   \mu +    \right. & \nonumber \\
 + \left.  \sum_{n=1}^{\infty} e^{-i (\phi -\phi') \frac{n-\delta}{\alpha}}
          \int_{0}^{\infty} J_{\frac{n-\delta}{\alpha}}(r\mu)
          J_{\frac{n-\delta}{\alpha}} (r'\mu) e^{-t\mu^{2}} \, \mu d
          \mu   \right\}  &  \label{green3}
\end{eqnarray}
where $\delta=\alpha ~{\rm mod}~1 $ and is valid for Re $\nu > -1$
($\nu$ is the index of the Bessel function).
For the heat kernel of the operator $L L^{\dag}$ the same
representation holds with $\delta= 2\alpha~{\rm mod}~1$.

One can compute the constant term $c^K_0$ in the asymptotic expansion of the
heat kernel (\ref{green3}) by a well know procedure \cite{dow} to
obtain for $K$
\begin{equation}
\displaystyle c^K_0= \frac{\delta(\delta - 1)}{2\alpha} +
\frac{1-\alpha^{2}}{12\alpha }.
\label{dow}
\end{equation}

 On the other hand the self adjoint extensions of $L^{\dag}L$ and
$LL^{\dag}$ depend on the boundary conditions one imposes on the
eigenfunction at the singularities. The choice of Dowker and of Aurell
and Salomonson for the Laplace--Beltrami operator is Dirichlet
boundary conditions.  This is equivalent to imposing for the
phase-shift $\delta$ in eq.(\ref{green3}) the restriction $0<\delta
\leq 1$.  This gives for small angular deficits i.e. $\alpha=1 +
\varepsilon$, for $\varepsilon < 0$, $\displaystyle c^K_0 =
\frac{\varepsilon}{3} + O(\varepsilon^{2})$ and $\displaystyle c^H_0 =
\frac{5\varepsilon}{6} + O(\varepsilon^{2})$ while for
$\varepsilon>0$, $\displaystyle c^K_0 = -\frac{2\varepsilon}{3} +
O(\varepsilon^{2})$ and $\displaystyle c^H_0 = -{7\over 6}\varepsilon
+ O(\varepsilon^{2})$.

Such a result is non analytic in $\alpha$ near the flat space and it
gives  the wrong continuum limit ($\varepsilon \rightarrow 0$)
for the FP determinant, both for positive and negative $\varepsilon$.

The reason of such a failure can be understood as follows: Dirichlet
boundary conditions are equivalent to cutting off the tip of the cone;
on the other hand the meaning of the tip of the cone of the Regge
surface is that of a locus of infinite curvature. Thus we looked at
the problem of treating the cone as the limit case of a regular
geometry:  for positive curvature we
described the tip of the cone as a segment of a sphere which connects
smoothly with the cone and for negative curvature we described
the tip of the cone as a segment of the Poincar\'e pseudo-sphere of
constant negative curvature.

The limit we are interested in, is the one of the radius of the sphere
going to zero, keeping constant the integrated curvature.  The
sphere of radius ${1\over 2}\rho$, of constant curvature $R=-2
e^{-2\sigma} \Delta \sigma = 8\rho^{-2}$ or the pseudo-sphere of constant
curvature $R=-8\rho^{-2}$ are described on
the complex $\omega$ plane by the conformal factor
$e^{2\sigma}=(1\pm u \bar u)^{-2}$ with $u=\omega/\rho$.
In order to proceed one needs to solve the eigenvalue problem on such
regularized cones.

Solving explicitly the eigenvalue equation we find, for the eigensolutions
with orbital angular momentum $m$ on the sphere
\begin{eqnarray}
m=n\geq0 ~~~~~ & \xi^{(n)} =
\displaystyle{\frac{u^{n}}{(1+u\bar{u})^{2}}}  ~_{2}F_{1}
(\gamma_{1} + 2, 1 - \gamma_{1}; n + 1 ; \frac{u\bar{u}}{1+ u\bar{u}}) \\
m=-n\leq0 &  \xi^{(n)} = \bar{u}^{n} ~_{2}F_{1} (\gamma_{1}, -1
- \gamma_{1}; n+1; \displaystyle{\frac{u\bar{u}}{1+ u\bar{u}}})
\end{eqnarray}
where $\gamma_{1} = \frac{1}{2}(-1+\sqrt{9+4(\rho\mu)^{2}}~)$.

Similar solutions are found in the case of the pseudosphere. Such
solutions have to be matched to the exterior solution on the cone
by imposing the continuity of the
logarithmic derivative of $e^{-2\sigma} \xi$ with respect to
$\bar{\omega}$ at $\mid \! \omega\!\! \mid = \tau_{0}$ (being
$\tau_{0}$ the radius on the $\omega$--plane at which the sphere
connects to the cone) as required by the structure of eigenvalue
equation $e^{-2\sigma} \frac{\partial}{\partial\omega} e^{2\sigma}
\frac{\partial}{\partial\bar{\omega}} e^{- 2\sigma} \xi = - \mu^{2}
\xi$.  The general eigensolution on the exterior cone for orbital
angular momentum $m$ has the form
\begin{equation}
\xi_{\rm ext}^{(m)} = \left( \frac{u}{\bar{u}} \right)^{\frac{m}{2}} (u
  \bar{u})^{\frac{\alpha -1}{2}} \left[ a(\rho) J_{\gamma}
(2\rho\mu p (u\bar{u})^{\frac{\alpha}{2}})  + b(\rho) J_{-\gamma}
  (2\rho\mu p (u\bar{u})^{\frac{\alpha}{2}}) \right]
\end{equation}
where $\gamma=\frac{m+\alpha -1}{\alpha}$ and  $p
=\frac{(u_{0}\bar{u}_{0})^{\frac{1-\alpha}{2}}}{\alpha(1 \pm
u_{0}\bar{u}_{0})}$. The coefficients $a(\rho)$ and $b(\rho)$ are fixed by
imposing the matching conditions at $\tau_{0}$. Letting $\rho
\rightarrow 0$ (and thus $\tau_{0} \rightarrow 0$) gives the result
\cite{pmppp}
that for the opening of the cone $2\pi\alpha$ with $\frac{1}{2} <
\alpha < 2$, only the term $J_{\frac{m+\alpha -1}{\alpha}}$ survives
for $m\geq 0$, while for $m<0$ the surviving term is
$J_{-\frac{m+\alpha -1}{\alpha}}$. Going over to the coordinate $z$,
the heat kernel $K(x,x';t)$ is thus given by (\ref{green3}) with
$\delta = \alpha -1$.

We come now to the heat kernel $H$ for the field $h$. The
requirement \cite{alva} that $\det'(L^{\dag}L) = \det'(LL^{\dag})$
fixes the eigenfunctions of $LL^{\dag}$ to $h=L\xi$. Thus the
eigenfunctions of $LL^{\dag}$ are given in the $z$ coordinate for
$m\geq0$ by
\begin{equation}
  \left. \frac{\partial}{\partial\bar{z}} \left[ \left( \frac{z}{\bar{z}}
\right)^{\frac{\gamma}{2}} J_{\gamma} (2 \mu (z \bar{z})^{\frac{1}{2}})
\right] \right.
\end{equation}
which through a well known identity on the Bessel functions equals
\begin{equation}
  - \mu \left( \frac{z}{\bar{z}} \right)^{\frac{\gamma + 1}{2}}
  J_{\gamma + 1} (2 \mu (z \bar{z})^{\frac{1}{2}}),
\end{equation}
while for $m<0$ they are given by
\begin{equation}
  \mu \left( \frac{z}{\bar{z}} \right)^{\frac{\gamma + 1}{2}}
  J_{-\gamma - 1} (2 \mu (z \bar{z})^{\frac{1}{2}}),
 \end{equation}
always with $\displaystyle{\gamma = \frac{m + \alpha - 1}{\alpha}}$.
The net result is that the heat kernel $H$ is given by (\ref{green3})
with $\delta = 2\alpha -1$.

Applying Dowker's procedure to $ K =K_{\alpha,\alpha-1}$,
eq.(\ref{green3}), we obtain
\begin{equation}
  c_{0}^{K} =
  \frac{(\alpha - 1)(\alpha - 2)}{2\alpha} + \frac{1 - \alpha^{2}}{12 \alpha}
\label{zeta1}
\end{equation}
and for $H = K_{\alpha, 2 \alpha -1}$
\begin{equation}
   c_{0}^{H}=
  \frac{(2\alpha - 1)(2\alpha-2)}{2\alpha} + \frac{1 - \alpha^{2}}{12 \alpha}
\label{zeta2}.
\end{equation}
We notice that the $c_{0}$ are analytic in $\alpha_{i}$ and $2(c_{0\
i}^{K} - c_{0\ i}^{H}) = 3 (1-\alpha_{i})$.  This holds for the
contribution to $Z(0) + {\rm dim~ker}L^{\dag}L$ of a single conical
singularity. Thus for a generic compact surface without boundary due
to the local nature of the coefficients of the asymptotic expansion
of the trace of the heat kernel \cite{dow} such a relation becomes $ 2
(c_{0}^{K} - c_{0}^{H}) = 3\sum_i(1-\alpha_i) = 3\chi$ where the sum
runs over the vertices and $\chi$ is the Euler characteristic of the
surface, in agreement with the Riemann-Roch index
theorem applied to $L^{\dag}L$ and $LL^{\dag}$ \cite{alva}.
This provides an interesting check of the consistency of our
regularization procedure.

We remark that the obtained results on the behavior of the
eigenfunctions at the origin in the limit when the regulator goes to
zero are largely independent of the details of the regularization of
the tip of the cone. In fact the
effect of our regularization is that to impose (apart from a
correction that behaves like $\rho^{2}$ and that vanishes in the limit
$\rho \rightarrow 0$) a fixed logarithmic derivative of $e^{-2\sigma}
\xi$ at the boundary, combined with the fact that for $m\geq 0$ the
regular eigenfunction of $L^{\dag}L$ to the null eigenvalue has the
form $e^{2\sigma} \omega^{m}$. The advantage of the spherical
regularization is to allow and explicit calculation of the
regularized eigenfunctions.

Imposing Dirichlet boundary condition on the field $\xi$
on a small circle and then making the radius of the circle vanish,
reproduces eq.(\ref{zeta1}) and eq.(\ref{zeta2}) only for $1 < \alpha <2$.
At the same time the field $h=\frac{\partial}{\partial \bar{z}} \xi$
will violate the Dirichlet boundary condition in the same range of
$\alpha$, as $\delta = 2 \alpha - 1$ no longer lies in the limits $0<
\delta \leq 1$.
With Neumann boundary conditions ($\frac{\partial}{\partial \bar{z}}
\xi  = 0$) we have the same situation in the interval $\frac{1}{2} <
\alpha <1$.

Our boundary conditions span the whole range $\frac{1}{2} < \alpha <2$, at
the boundary of which both the $L^{2}$ character of the eigenfunctions
and the procedure for obtaining eq.(\ref{dow}), for $\delta =
\alpha - 1$ or $\delta = 2\alpha -1$, are lost.

For $0<\alpha \leq \frac{1}{2}$ with our boundary conditions we obtain
$\delta = \alpha$ and for $N < \alpha \leq N+1$ we have $\delta =
\alpha -N$, thus introducing a non analytic behavior of ${\det}' L^{\dag}L$
as a function of $\alpha$.
This is not completely unexpected as a non analytic behavior is
already present in $c_{0}$ for the Green function of a particle on a
cone as a function of the magnetic flux through the tip of the cone
\cite{dow}, which in our case represents the phase change. The fact that
the analytic continuation of the eigenfunctions outside the range
$1/2<\alpha<2$ are no longer $L^2$ integrable suggests that an
analytic extension of eq.(\ref{zeta1}) and eq.(\ref{zeta2}) to the whole range
$0<\alpha$ requires a definition of the determinant of a
transformation on non $L^2$ functions.

Coming back to eq.(\ref{zeta}) and performing a variation of the
conformal factor we obtain
\begin{eqnarray}
\lefteqn{ - \delta \ln \frac{{\det}'(L^{\dag}L)}{\det(\Psi_{i},
    \Psi_{j})} = \gamma_{E} \delta c_{0}^{K} + \sum_{i} \left\{
    (\delta \lambda_{i} -\lambda_{i} \frac{\delta
    \alpha_{i}}{\alpha_{i}}) [ 4 c_{0\ i}^{K} - 2 c_{0\ i}^{H}] +
    \right. } & \\
& + \left. {\mbox{Finite}}_{\epsilon \rightarrow 0} \left[ 4
    {\displaystyle\frac{\delta \alpha_{i}}{\alpha_{i}}} \int\!dx\, \ln
    (\alpha_{i}|x|) \, K_{\alpha_{i}}(x,x,\epsilon) -2
    {\displaystyle\frac{\delta \alpha_{i}}{\alpha_{i}}} \int\!dx\, \ln
    (\alpha_{i}|x|) \, H_{\alpha_{i}}(x,x,\epsilon) \right] \right\}.
    \nonumber
\end{eqnarray}
A differential of this structure \cite{aur} can be integrated to give
\begin{eqnarray}
\lefteqn{\ln \sqrt{\frac{{\det}'(L^{\dag}L)}{\det(\Psi_{i},
\Psi_{j})}} = } \label{fin} \\
& {\displaystyle
=\frac{26}{12} \left\{ \sum_{i,j\neq i}
\frac{(1-\alpha_{i})(1-\alpha_{j})}{\alpha_{i}} \ln |w_{i} - w_{j}| +
\lambda_{0} \sum_{i} (\alpha_{i} - \frac{1}{\alpha_{i}}) -
\sum_{i} F(\alpha_{i}) \right\}  }   \nonumber
\end{eqnarray}
where $F(\alpha)$ is given by a well defined and convergent
integral representation.

In the continuum limit, i.e.\ small angular deficits
$1-\alpha_{i}$ and dense set of $\omega_{i}$, the first two terms of
(\ref{fin}) go over to the well know continuum formula
\begin{equation}
{\displaystyle
\frac{26}{96\pi} \left\{\int  dx\,dy \;(\sqrt{g} R)_{x}
\frac{1}{\Box}(x,y)(\sqrt{g} R)_{y}
-2 (\ln \frac{A}{\bar{A}})
\int \: dx\,\sqrt{g} R \right\} }  \label{liouv2}
\end{equation}
as can be easily checked, where $A$ is the area $\int dx\,\sqrt{g}$
and $\bar{A}$ is the area evaluated for $\lambda_{0}=0$. The
remainder $\sum_{i}
F(\alpha_{i})$ goes over to a constant topological term.

In addition to having the correct continuum limit eq.(\ref{fin}) has the
following appealing features:

i) It is an exact result giving the
value of the $F.P.$ determinant related to the conformal gauge-fixing
on a two dimensional Regge surface.

ii) It is invariant under the group $SL(2,C)$ which acts on
$\omega_{i}$ and $\lambda_{0}$ as

\begin{eqnarray}
  \lefteqn{\omega_{i} \rightarrow \omega_{i}' =
  \frac{a\omega_{i} +b}{c\omega_{i} + d}} \\
  & \lambda_{0} \rightarrow \lambda_{0}' = \lambda_{0} + \sum_{i}
  (\alpha_{i} -1) \ln |\omega_{i} c + d| \nonumber
\end{eqnarray}
and leaves the $\alpha_{i}$ unchanged.

iii) While $\alpha_{i}>0$ with $\sum_{i}(1-\alpha_{i})=2$, the
  $\omega_{i}$ vary without restriction in the complex plane.
  As pointed out in \cite{foer} this is an advantage over the
  equivalent parameterization of the Regge surface in term of the bone
  lengths $l_{i}$ where one has to keep into account of a large number
  of triangular inequalities.

\end{document}